# Remarks on the acceleration of global warming and the imminent breach of the 1.5°C Paris Agreement target

Erhard Reschenhofer[1]

**Abstract**

To answer the questions of whether global warming is accelerating and when the 1.5°C Paris Agreement target will be exceeded, the global mean surface temperature from 1880 to 2025 is first examined using a purely graphical approach and later, in a more conventional way, using various time-domain and frequency-domain methods. In an effort to reduce variability, exogenous variables such as El Niño and solar variations are taken into account. Although it ultimately remains unclear to what extent these variables are actually helpful, we feel confident in summarizing the empirical results of this study to suggest that global warming is indeed accelerating and that a breach of the 1.5°C Paris Agreement target is imminent. But when it comes to statistical significance, caution should still be exercised. While the acceleration hypothesis can be confirmed with a fair degree of certainty under reasonably plausible assumptions (albeit with the help of a bit of data snooping, which is unavoidable when building on the results of earlier studies that used virtually the same data), there is currently not enough evidence to prove that the 1.5°C target has already been exceeded. However, if 2026 and 2027 turn out to be very warm due to the approaching El Niño, that could change very soon.

**Keywords:** global warming, 1.5°C Paris Agreement threshold, warming surge, structural breaks, El Niño, sunspot number, Atlantic Multidecadal Oscillation.

## 1. Introduction

While there is broad scientific consensus that man-made global warming is happening (see, e.g., Powell, 2019; Lynas and Perry, 2021), two important research questions are still being debated. The first is whether global warming is accelerating (Foster and Rahmstorf, 2011; Risbey et al., 2018; Jenkins, Sanderson et al., 2022; Jenkins, Povey et al., 2022; Richardson, 2022; Samset et al., 2023; Beaulieu et al., 2024; Hansen et al., 2025; Foster and Rahmstorf, 2026), and the second is when the increase in global mean surface temperature (GMST) compared to pre-industrial levels will reach the critical threshold of 1.5 °C (Bevacqua et al., 2025; Cannon, 2025; Reschenhofer, 2025).

In order to answer the first question as objectively as possible, Beaulieu et al. (2024) fitted piecewise linear regression models to different GMST datasets, but found no statistically significant (at the 5% level) warming surge since the 1970s once the implausible assumption of no autocorrelation was discarded. However, when Foster and Rahmstorf (2026) attempted to reduce the noise level by taking the spurious effects of exogenous factors such as El Niño, volcanism, and solar variations into account, they apparently were able to detect a significant acceleration since around 2015. In general, including additional variables in a model makes sense only if the information they provide outweighs the additional uncertainty caused by errors in measuring these variables, estimating the additional model parameters, and specifying the larger model. The danger that a barely significant result obtained using a complex model with many parameters and meta-parameters is not genuine but just due to data snooping is very real.

---

[1] *Retired from University of Vienna, E-mail: erhard.reschenhofer@univie.ac.at*

The model of Foster and Rahmstorf (2026) describes the monthly GMST as the sum of a deterministic trend, three additive terms representing the exogenous factors, and a stationary error process. Of course, it is still necessary to specify the deterministic trend in more detail (e.g., by the number of breaks in the case of a broken linear trend or by the highest degree in the case of a polynomial trend), to choose suitable lags for the exogenous factors, and to select an ARMA order for the error process, before any estimating or testing can be carried out. Since each of these tasks is extremely complex and the next steps (e.g., identifying the breakpoints, calculating critical values) are not trivial either, it is hard to imagine that the whole procedure could produce undisputable results. Moreover, the validity of such a model would be seriously called into question if there were any seasonal fluctuations, which would clearly be inconsistent with stationarity. From a naive point of view, seasonality cannot occur at all because monthly temperatures are usually expressed as anomalies which represent the departure of a specific month's temperature from a long-term average for the same month. Unfortunately, seasonality still occurs in the anomalies because different months are affected by global warming in different ways (see Reschenhofer, 2024). There is no simple procedure to deal with time-varying seasonality. However, adding one or two sinusoids with fixed parameters (as in Foster and Rahmstorf, 2011) might work if the focus is on a reasonably stable subperiod.

Evidence of an accelerated rise in temperature would also have implications for the second research question regarding the 1.5°C Paris target. In fact, the standard procedure for measuring the temperature rise would no longer make sense in the event of an acceleration. Namely, if we first determine the baseline temperature by calculating the average over the pre-industrial period and then track subsequent trends using rolling averages calculated over periods of 2 or 3 decades, we tacitly accept a considerable delay, but in return expect a high degree of accuracy. Unfortunately, this expectation cannot be met if the temperature rise accelerates. Not only would an exceedance of the threshold not be noticed until many years later, but the estimated time of the breach would also be highly inaccurate. When there is a significant deviation from a linear trend, simple symmetric averages are highly biased and therefore no longer appropriate. The most recent 241-month moving average that we can calculate is for the period from December 2005 to December 2025. However, due to a possible break in the trend around 2010, this value cannot be used as an estimate of the trend in December 2015. It would significantly overestimate the true value.

So are there any alternatives now, if complex models based on implausible assumptions are unreliable and simple averages are severely distorted? In the next section, we take a purely graphical approach and let the data speak for themselves. Of course, we do not limit ourselves to raw data alone. Straightforward deterministic data transformations are still permitted, but data-driven transformations are not. Section 3 takes a somewhat more orthodox approach. Time series analyses are conducted both in the time domain and in the frequency domain. The results are largely consistent with those of the graphical analysis in the previous chapter. However, given the uncertainties mentioned above, any "statistically significant" results must be interpreted with the utmost caution. Section 4 concludes.

## 2. Graphical analysis

For our analysis, we use the HadCRUT5 surface temperature dataset (Morice et al., 2021; https://crudata.uea.ac.uk/cru/data/temperature/), which contains global annual and monthly averages of the combined land and marine temperatures from January 1850 to December 2025. The empirical analysis is performed using the statistical software R (R Core Team, 2024). One way to describe how GMST changes over time is to fit a simple deterministic trend. Figure 1.a shows broken linear trends fitted by OLS and GLS, respectively. Although the deviations from the trends are highly autocorrelated, the simple OLS procedure produces practically the same result as the more sophisticated GLS procedure, which allows for autocorrelation. In contrast, the choice of the estimation period does seem to make a difference. Due to concerns about data quality, we have omitted the first 30 years, resulting in a baseline that is significantly below the average for the period from 1850 to 1900, which is commonly used as a proxy for pre-industrial levels. Clearly, the baseline plays a major role in determining when a temperature increase of 1.5 degrees Celsius compared to pre-industrial levels will be reached.

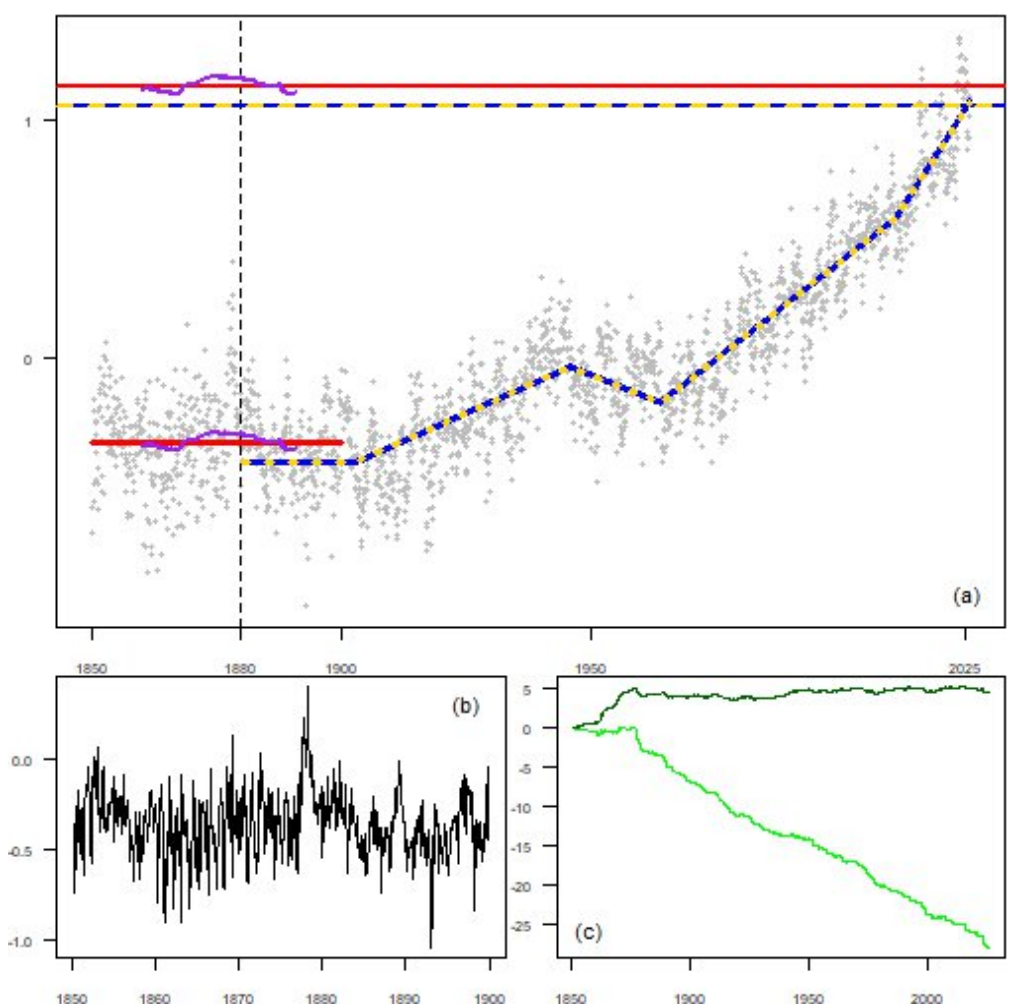

Figure 1: (a) Broken linear trend with four structural breaks (at 1903.3, 1945.11, 1963.12, and 2010.11) fitted by OLS (blue) and GLS (yellow) to the monthly GMST (HadCRUT). The horizontal lines at the top represent increases of 1.5°C compared to the average for the period from 1850 to 1900 (red) and the initial level of the trend, respectively. The purple lines are moving averages over 241-month periods. (b) Small segment of the monthly GMST. (c) Estimation of first autocorrelation of 12th differences with cumulative sums (1): It was approximately 0.25 before 1880 (green line is flat for $c = 0.25$) and much greater after 1880 (decrease in green line). It was approximately 0.65 after 1880 (dark green line is flat for $c = 0.65$) and much smaller before 1880 (increase in dark green line).

Our skepticism regarding measurements taken before 1880 stems from significant differences compared to later measurements (see Figure 1.b). The first-order autocorrelation of the 12th differences was approximately 0.25 before 1880 and 0.65 after 1880 (see Figure 1.c). This discrepancy is far too large to be attributed solely to random fluctuations. To track the autocorrelation over time, the procedure proposed by Reschenhofer (2024) was used. It is based on the cumulative sums

$$\sum_{t=2}^{j} \left( c v_t^2 - v_t v_{t-1} \right), j = 2,\ldots,n, \qquad (1)$$

which are obtained from the time series $v_1,\ldots,v_n$ (in our case the 12th differences of the GMST) and plotted for different values of $c$. A decrease/increase in (1) suggests that the autocorrelation is grater/less than $c$.

To estimate the magnitude of the increase in temperature compared to pre-industrial levels, we have to make a series of more or less arbitrary decisions. Since the increase is the difference between an initial value and a final value, it naturally depends on both. As for the latter, we have already rejected the idea of a multi-decade average. An obvious alternative is the endpoint of a fitted trend, which, of course, depends heavily on the type of the trend (broken linear trend, polynomial trend or a nonparametric trend such as the Hodrick-Prescott filter) and the choice of certain tuning parameters (number of breaks, degree of polynomial, etc.). Determining the baseline value is not easy either. We could use the average value for a given period or the starting value of a fitted trend. In either case, the question arises whether we should exclude some of the data due to potential quality issues.

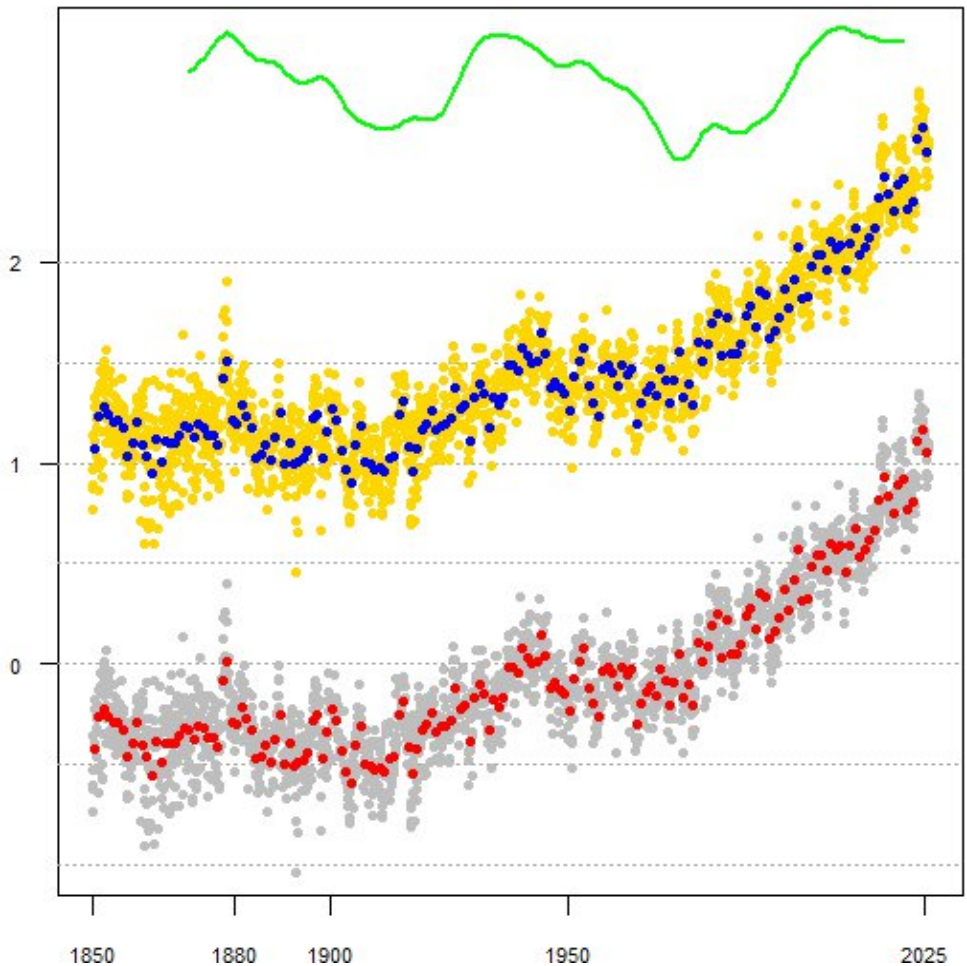

Figure 2: Comparison of the monthly (gray) and annual (red) GMST (HadCRUT) with GMST shifted upward by 1.5°C (yellow and blue, respectively). The green line shows the annual AMO from 1870 to 2020 (MOHC).

Fortunately, we can avoid all these problems just by looking at Figure 2, which compares the raw data with the raw data shifted upward by 1.5°C. Apparently, the last three years of the original data already fall within the narrower range of variation of the shifted pre-industrial period (regardless of how exactly one defines this period), which suggests that a breach of the 1.5°C target is imminent. But we definitely need a few more monthly highs before we can start thinking about statistical significance. With the next El Niño on the horizon, that might even happen as early as 2027. This would be consistent with an earlier paper (Reschenhofer, 2025), which, based on simulations, also predicted a breach before 2028.

Although it is unlikely that there is a strong correlation between the global climate and a local phenomenon such as the Atlantic Multidecadal Oscillation (AMO), an AMO graph is still shown in Figure 2 because of certain similarities. The annual AMO was downloaded from the Climate Dashboard (https://climate.metoffice.cloud/) of the Met Office Hadley Centre (MOHC). However, if global temperatures do indeed exhibit a long-term oscillatory component similar to the AMO, this oscillation would currently be on the decline and would slow the warming somewhat. Or, to put it less optimistically, without this effect, the current rapid warming would be even more pronounced.

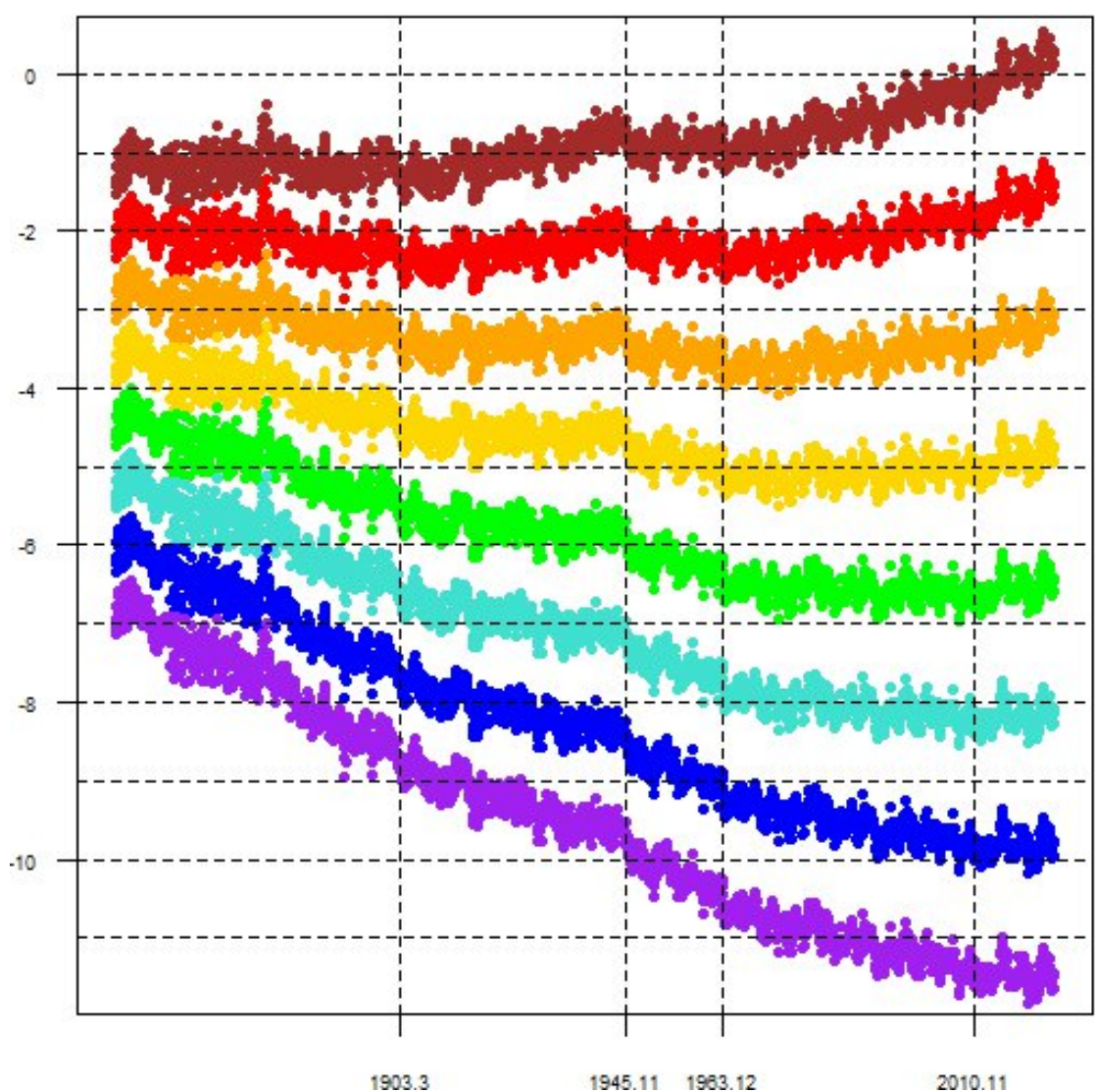

Figure 3: GMST (HadCRUT) minus a trend with a decline of 0°C (brown), 0.05°C (red), 0.1°C (orange), 0.15°C (yellow), 0.2°C (green), 0.25°C (turquoise), 0.3°C (blue), and 0.35°C" (purple) per decade.

When we turn to the question of whether global warming might be accelerating, we find a positive answer surprisingly easily and quickly. Figure 3 shows what happens when linear trends with different slopes are subtracted from the temperature data. Only when we subtract the "correct" linear trend does the trend in the temperature data disappear. Possible candidates for periods in which there is no apparent trend include the brown cluster before 1903.3 and between 1945.11 and 1963.12, the yellow cluster between 1903.3 and 1945.11, the green cluster between 1963.12 and 2010.11, and the turquoise cluster since 2010.11. The colors brown, yellow, green, and turquoise represent increases of 0°C, 0.15°C, 0.2°C, and 0.25°C per decade, respectively. We conclude from this that global warming accelerated from approximately 0.2°C to 0.25°C per decade about 15 years ago.

## 3. Time series analysis

Any seasonal variation that may exist must be taken into account in every advanced analysis or modeling exercise, because it is incompatible with the standard assumption of stationary errors. Although the use of anomalies should actually rule this out, there are still clear signs of seasonal patterns in the periodograms of various subseries (detrended if necessary) of the GMST time series (see Figure 4). This is true for the period from 1880 to 2025 as well as for the subperiods 1850-1903.3, 1880-1903.3, and 1903.4-1963. The seasonal pattern is most noticeable in the subperiod 1850-1903.3, which also includes the questionable data. In the subperiod 1903.4-1963, it is already much less pronounced and also deviates significantly from the sine waveform, as evidenced by the small size of the periodogram ordinate at the fundamental frequency $\pi/6$ compared to the ordinates at the higher-order harmonics. It is pure luck that just the subperiod 1964-2025, which we need for the investigation of the acceleration hypothesis, happens to exhibit only very weak seasonality, so that we could even get away with ignoring it entirely.

Since we have just used periodograms and will also be performing cross spectral analysis later on, let us summarize a few important definitions here. The cross covariance function $\gamma_{xy}$, the cross spectrum $f_{xy}$, the amplitude spectrum $R_{xy}(\omega)$, the squared coherency $\rho_{xy}^2(\omega)$, and the phase spectrum $\varphi_{xy}(\omega)$ of a bivariate stationary process with components $(x_t)$ and $(y_t)$ are defined by

$$\gamma_{xy}(j) = \text{cov}(x_t, y_{t-j}), \tag{2}$$

$$f_{xy}(\omega) = R_{xy}(\omega) e^{i\varphi_{xy}(\omega)} = \frac{1}{2\pi}\sum_{j=-\infty}^{\infty} e^{-i\omega j}\gamma_{xy}(j), \tag{3}$$

and

$$\rho_{xy}^2(\omega) = \frac{R_{xy}^2(\omega)}{f_{xx}(\omega)f_{yy}(\omega)} = \frac{\left|f_{xy}(\omega)\right|^2}{f_{xx}(\omega)f_{yy}(\omega)}, \tag{4}$$

respectively. Their sample counterparts are obtained by replacing the theoretical covariances $\gamma_{xy}(j)$ by the sample covariances

$$\hat{\gamma}_{xy}(j) = \begin{cases} \frac{1}{n}\sum_{t=j+1}^{n}(x_t-\bar{x})(y_{t-j}-\bar{y}), & j=0,\ldots,n-1, \\ \frac{1}{n}\sum_{t=1}^{n-j}(x_t-\bar{x})(y_{t-j}-\bar{y}), & j=-1,\ldots,-(n-1). \end{cases} \tag{5}$$

In the univariate case, the autocovariance function, the spectral density, and the periodogram are defined by $\gamma_{yy}$, $f_{yy}$, and $\hat{f}_{yy}$, respectively. Although other frequencies may be of interest for specific questions (see Reschenhofer and Mangat, 2021), we will follow the standard procedure and evaluate the periodogram $\hat{f}_{yy}$ as well as the other sample functions defined in the frequency-domain only at the Fourier frequencies $\omega_k = 2\pi k/n$, $k = 1,...,[n/2]$.

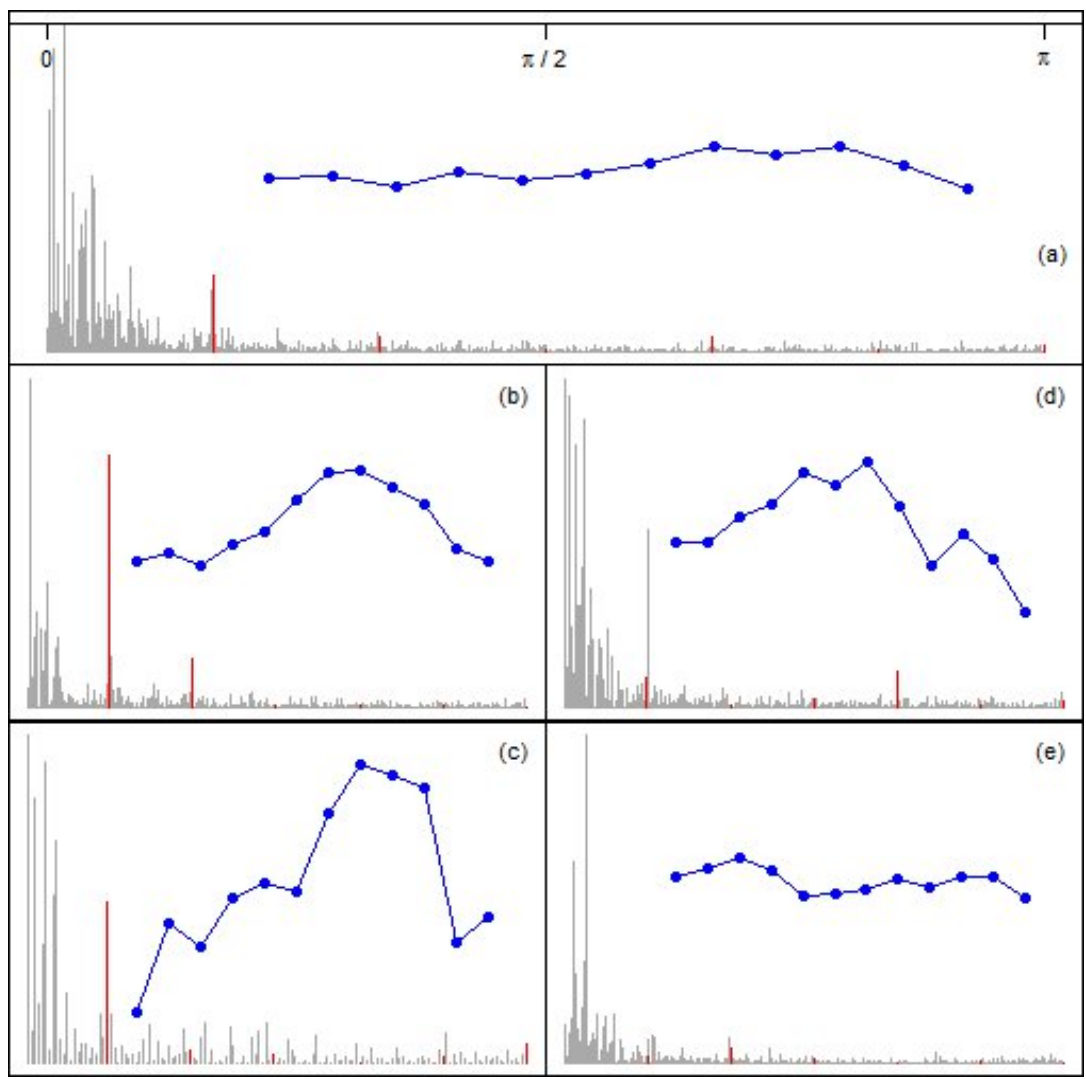

Figure 4: Periodograms (dark gray, red at fundamental frequency $\pi/6$ and at higher-order harmonics) of detrended (if necessary) monthly GMST (HadCRUT) time series and seasonal patterns (blue): (a) 1880-2025 (b) 1850-1903.3 (c) 1880-1903.3 (d) 1903.4-1963 (e) 1964-2025

Apart from the seasonal patterns, the periodograms in Figure 4 show high power in the low-frequency region indicating a strong positive autocorrelation. This autocorrelation must, of course, be taken into account when assessing the value of exogenous variables for forecasting. It often turns out that lagged exogenous variables no longer contribute as much once the own history is properly accounted for. The primary candidates for exogenous factors influencing global temperature are solar activity and El Niño. Monthly means of the daily total sunspot numbers are available since 1749. The data were obtained (Licence: CC BY-NC: https://creativecommons.org/licenses/by-nc/4.0) from the World Data Center SILSO (WDC-SILSO), Royal Observatory of Belgium (https://doi.org/10.24414/qnza-ac80, https://www.sidc.be/SILSO/datafiles). As an indicator for the El Niño, the Oceanic Niño Index (ONI) was downloaded from the Climate Prediction Center (CPC; https://www.cpc.ncep.noaa.gov/). The ONI (available from 1950 to 2025) is the rolling 3-month average of sea-surface temperature anomalies (relative to 1961-1990) for an area in the east tropical Pacific.

Fitting ARMA models

$$y_t = \phi_1 y_{t-1} + \ldots + \phi_p y_{t-p} + u_t + \theta_1 u_{t-1} + \ldots + \theta_q u_{t-q}, \qquad (6)$$

of order $(p, q)$ satisfying (i) $p \leq 30$, $q = 0$, (ii) $p, q \leq 4$, and (iii) $p, q \leq 6$ to each time series (1950 to 2025) and selecting the "best" order by minimizing the AIC statistic

$$n \log\left(\hat{\sigma}_u^2\right) + 2(p + q + 1), \qquad (7)$$

we obtain (27,0), (2,3), (6,6) for the mean sunspot number, (5,0), (4,3), (3,5) for the El Niño, and (4,0), (4,0), (6,5) for the detrended GMST. To verify that a model selected by AIC is indeed appropriate, we plotted an estimate of the ARMA spectral density

$$f(\omega) = \frac{\sigma^2}{2\pi} \left| \frac{1 + \theta_1 e^{-i\omega 1} \ldots + \theta_q e^{-i\omega q}}{1 - \phi_1 e^{-i\omega 1} \ldots - \phi_p e^{-i\omega p}} \right|^2, \qquad (8)$$

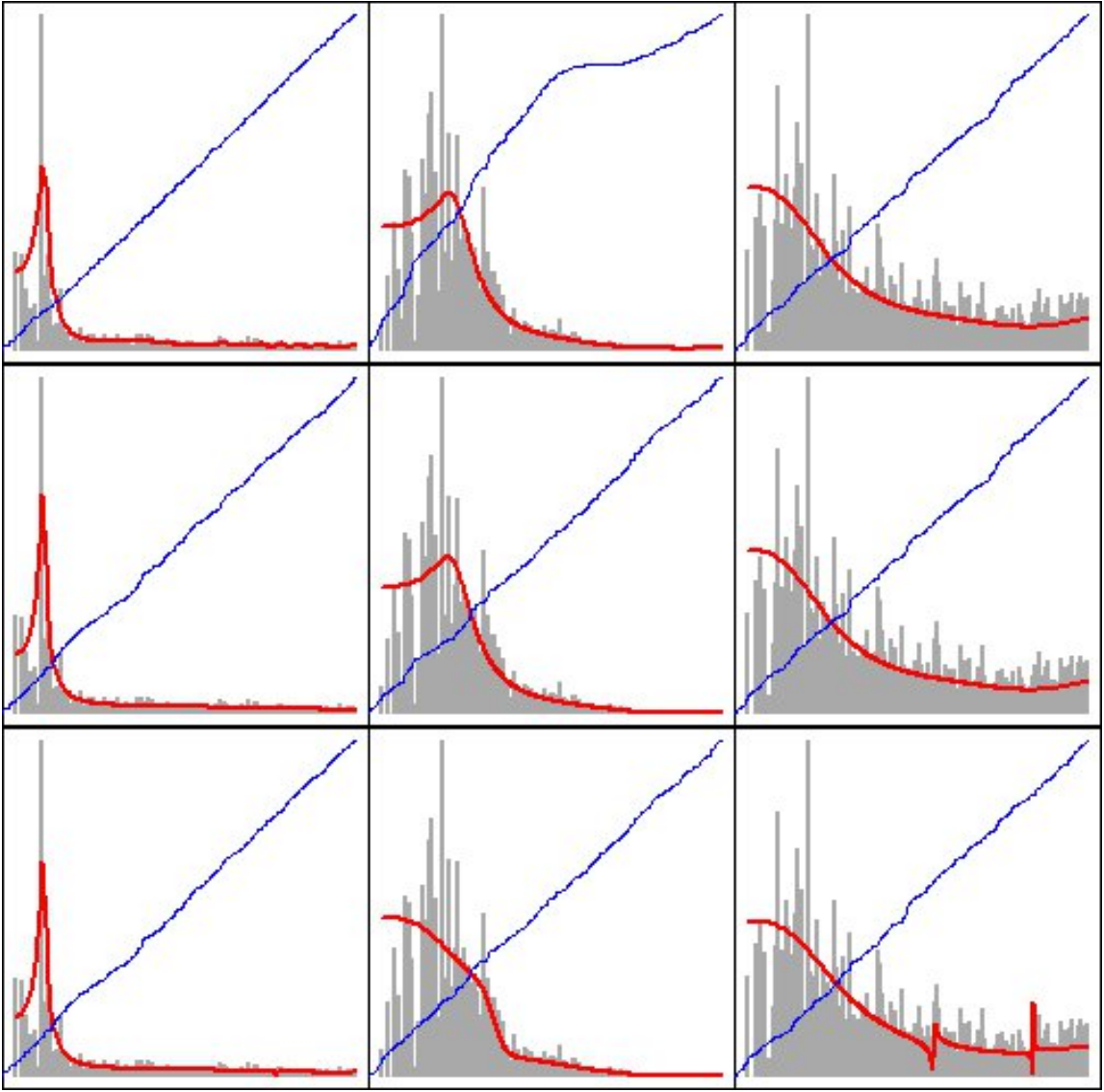

Figure 5: ARMA($p, q$)-spectral densities (red) fitted to periodograms (gray) of sunspot number (first column), El Niño (second column), and detrended GMST (third column) for the time period 1950-2025. AIC was used to select the best model satisfying $p \leq 30, q = 0$ (first row), $p, q \leq 4$ (second row), and $p, q \leq 6$ (third row). For better visibility, the periodograms and the spectral densities as well as the frequencies were subjected to a square root transformation. In contrast, linear scaling was used for the cumulative periodograms of the ARMA($p, q$) residuals (blue).

which represents the ARMA model in the frequency domain, alongside the periodogram, which represents the data in the frequency domain. In addition, we also checked the whiteness of the ARMA residuals by plotting their cumulative periodogram that should ideally resemble a straight line. Figure 5 shows that for the detrended GMST a satisfactory fit can already be achieved with a low-dimensional AR model. For the other two time series, either a high-dimensional AR model or, more parsimoniously, an ARMA model are needed. But more important anyway is the dynamics of the detrended GMST. We will keep in mind for later that in this case a simple AR(4) model seems reasonably plausible.

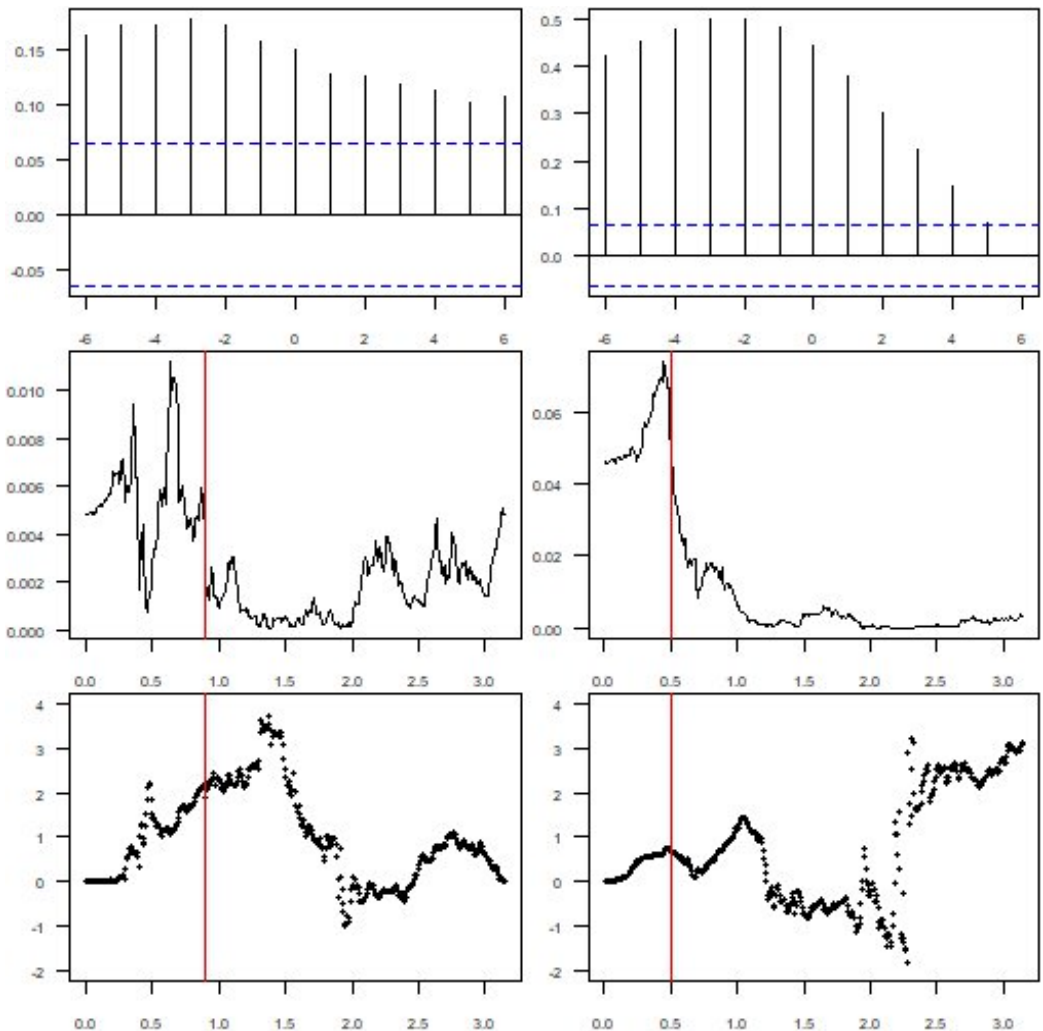

Figure 6: Using the sample cross-correlations (first row), the estimated squared coherency (second row), and the estimated phase (third row) for the analysis of the relationship in the period from 1950 to 2025 between the mean sunspot number (WDC-SILSO) and the detrended GMST (first column) and the ONI (CPC) and the detrended GMST (second column), respectively.

Before we move on to multivariate analysis, we would like to briefly explain the somewhat peculiar spectral density in the lower right corner of Figure 5. At first glance, the ARMA(6,5) spectral density selected by AIC seems to waste 8 parameters unnecessarily, because there are two conjugate pairs of roots of the numerator polynomial that almost cancel out with two conjugate pairs of roots of the denominator polynomial. Of course, that is not really the case. No optimization algorithm wastes parameters pointlessly if there is still so much variability in the periodogram that needs to be explained. This peculiar finding is simply a consequence of the fact that the smoothness of ARMA spectra makes it difficult to model a jump at a certain frequency.

Therefore, a sudden increase in the periodogram must be modeled by a zero at that frequency in the numerator polynomial immediately followed by a zero in the denominator polynomial.

A cross-spectral analysis is performed to examine the relationship between the mean sunspot number and the detrended GMST as well as the relationship between the El Niño and the detrended GMST. The squared coherency measures the strength of the linear relationship between the two components at different frequencies. The phase spectrum can be used to determine which component leads or lags at different frequencies. The amplitude spectrum, the squared coherency, and the phase spectrum can be estimated by smoothing their sample counterparts. Figure 6 shows the sample cross-correlations, the estimated squared coherency, and the estimated phase for each of the two relationships of interest. At the lowest frequencies, where the two squared coherencies are largest, the two phase spectra have approximately positive slopes of 2, which suggests that both the mean sunspot number and El Niño lead the detrended GMST by two months. Accordingly, the maximum cross-correlation is also reached at lag minus two or minus three.

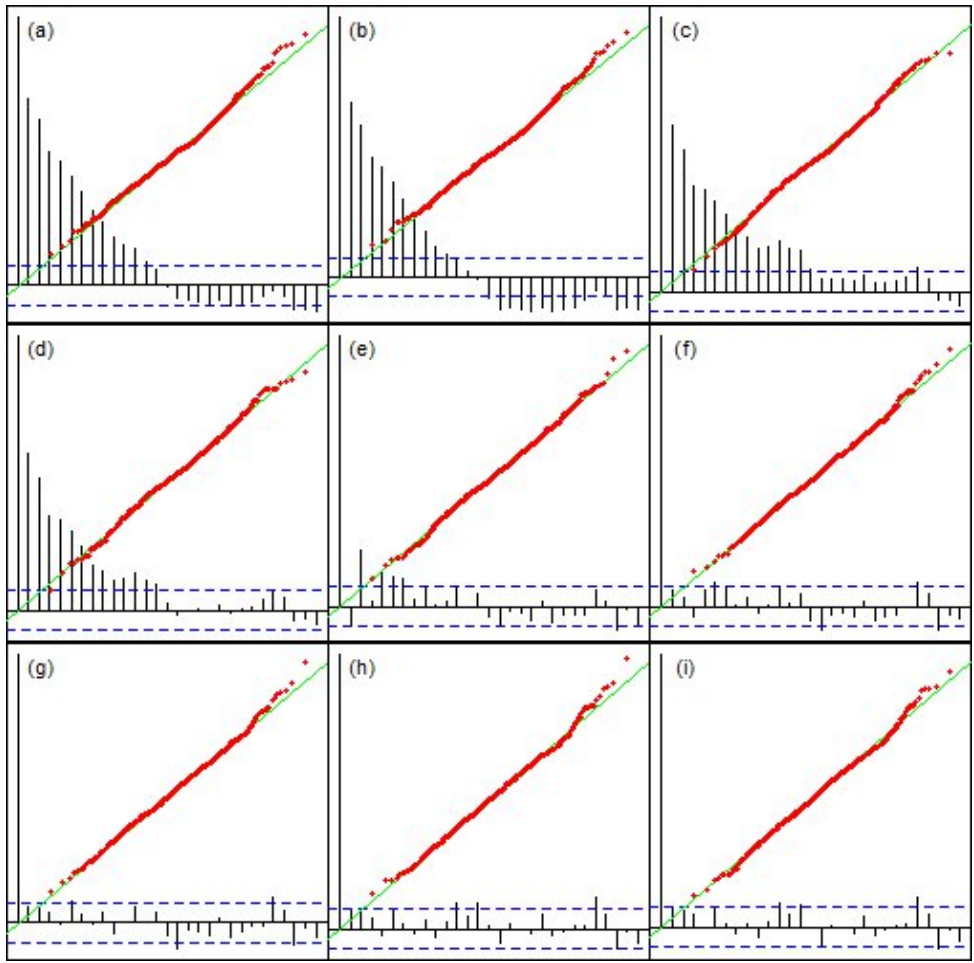

Figure 7: Sample autocorrelations and normal Q-Q plots (with 45° reference lines) of the standardized recursive residuals obtained by fitting the following models with the dependent variable $y$ (GMST) and the independent variables $x$ (mean sunspot number) and $z$ (El Niño):

(a) $y_t = a + bt + u_t$ (b) $y_t = a + bt + \beta_1 x_{t-2} + u_t$ (c) $y_t = a + bt + \beta_2 z_{t-2} + u_t$

(d) $y_t = a + bt + \beta_1 x_{t-2} + \beta_2 z_{t-2p} + u_t$ (e) $y_t = a + bt + \phi_1 y_{t-1} + u_t$

(f) $y_t = a + bt + \phi_1 y_{t-1} + \phi_2 y_{t-2} + u_t$ (g) $y_t = a + bt + \phi_1 y_{t-1} + \ldots + \phi_4 y_{t-4} + u_t$

(h) $y_t = a + bt + \phi_1 y_{t-1} + \ldots + \phi_4 y_{t-4} + \beta_2 z_{t-2p} + u_t$

(i) $y_t = a + bt + \phi_1 y_{t-1} + \ldots + \phi_4 y_{t-4} + \beta_1 x_{t-2} + \beta_2 z_{t-2p} + u_t$

Figure 7 compares various linear models with GMST as dependent variable and the mean sunspot number and El Niño as independent variables using sample auto-correlations and normal Q-Q plots of the standardized recursive residuals. While normality generally poses no problem, achieving near-whiteness requires at least two lags of GMST. This is also evident when plotting the cumulative absolute residuals. Once lags of GMST are included, the exogenous variables provide little additional benefit (see Figure 8.a). In particular, including the mean sunspot number never leads to a significant improvement, not even in the case of the pure trend model.

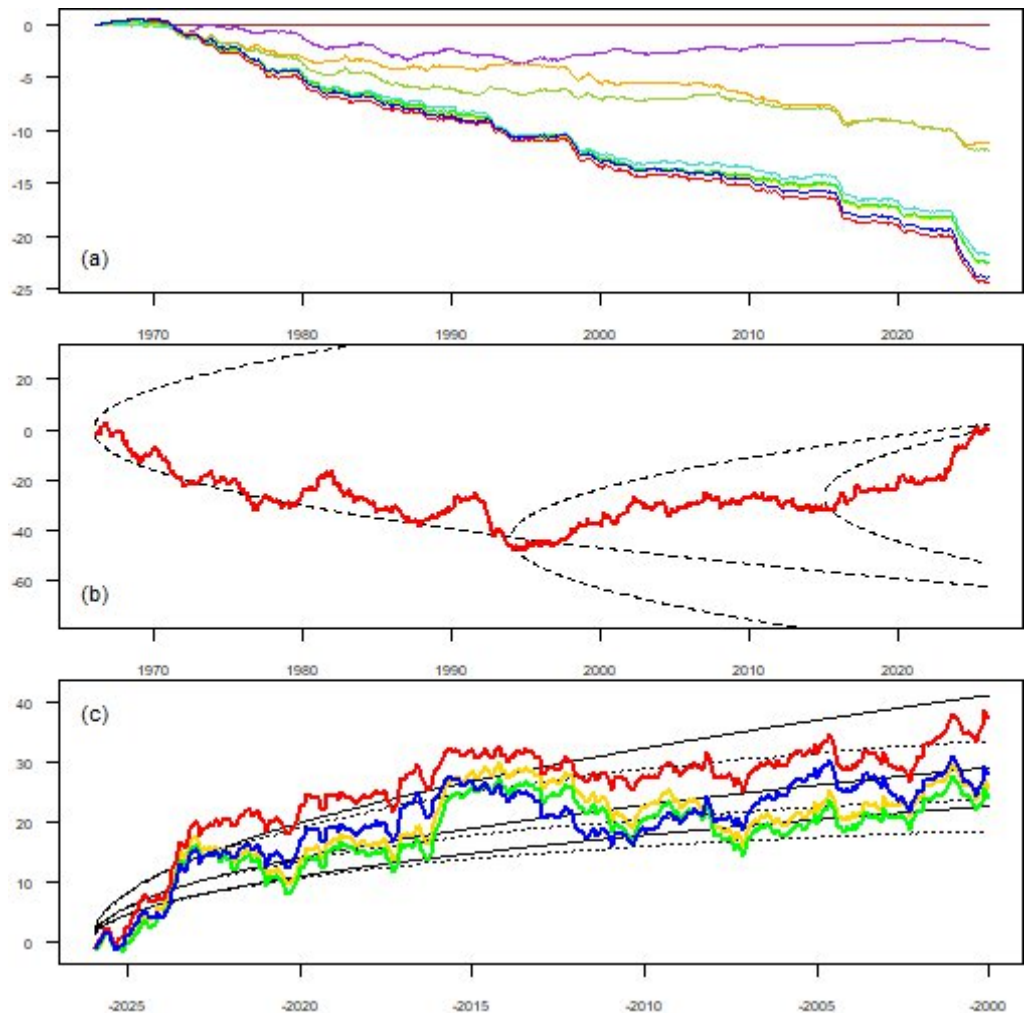

Figure 8: (a) The absolute recursive residuals of the nine models (a)-(i) described in Figure 7 and represented by the colors brown, purple, orange, yellow green, turquoise, yellow, green, red, blue are plotted cumulatively relative to the worst model (trend only).
(b) Cumulative standardized recursive residuals of the "best" model (h) with parabolas illustrating possible rejection regions at the one-sided 1% level.
(c) Cumulative standardized recursive residuals of models (f)-(i) from the present back to 2000. The parabolas defining the pointwise rejection regions for tests at the one-sided 10%, 5%, and 1% levels of significance were obtained either under idealized assumptions (solid lines) or using simulations (dotted lines).

If certain assumptions are met, such that the standardized recursive residuals are approximately i.i.d. $N(0,1)$, then it would be natural to use a sum of these residuals as a test statistic to detect a break in the trend. In the event of an acceleration, the residuals after the structural break would typically be positive. Ideally, the sum used as test statistic should include as few terms as possible before the structural break and as many terms as possible after it. However, the total number $k$ of terms included must not be too large, because the assumption of a $N(0,k)$ distribution for the test statistic under the null hypothesis becomes increasingly unrealistic as $k$ increases. In the extreme case, when all residuals are used, the sum is exactly zero due to standardization. Since the exact time of the break is usually unknown in practice, it might be tempting to increase the probability of rejecting the null hypothesis by data snooping (see Figure 8b). Strictly speaking, using a one-sided test instead of a two-sided one already constitutes data snooping. After all, it was only by examining the data (e.g., Jenkins, Povey et al., 2022) that we came up with the acceleration hypothesis in the first place. Why should a deceleration be ruled out from the outset?

Fortunately, there are a few things we can do to improve the situation a little. First, we may plot the cumulative residuals not from the beginning, but from the end, i.e., starting from the present. Furthermore, we can use rejection regions generated through simulations instead of the less reliable ones based on the $N(0,k)$ approximation. In both cases, our parabolas that define the rejection regions refer only to a single point in time. The advantage of this is that they are not supposed to hold simultaneously at many points and can therefore be narrower. The downside is that the number of residuals used for testing must be specified in advance, obviously without being allowed to look at the data. However, Figure 8.c shows that we always get a significant result at the one-sided 5% level for the largest model as long as we do not go back further than 2012. This restriction is not necessary for the 10% level. If the mean sunspot number is excluded, the 1% level can even be achieved. For the simulations, a linear trend model with stationary AR(4) errors was used. The model parameters were estimated over the relatively stable time period from 1964.1 to 2010.11.

## 4. Discussion

In empirical studies, it is common practice to provide evidence for a hypothesis of interest indirectly. Rather than seeking evidence in favor of the hypothesis, researchers look for evidence against the alternative. For our research questions, this means that we must find sufficient evidence to reject the hypotheses that "the 1.5°C Paris target has not been reached yet" and "the rate of warming is constant." Unfortunately, to achieve this, we must make a number of assumptions whose plausibility is not always beyond dispute. In the end, we cannot be sure whether a hypothesis has been rejected because it is false, or simply because some of the many technical assumptions are incorrect.

In the simplest case, we might assume that the GMST has followed a linear trend since the 1960s and that deviations from this trend can be described using a low-dimensional ARMA model. The rejection of the null hypothesis could then be interpreted as an indication of acceleration or deceleration provided that a broken linear trend were the only alternative to a linear trend. But of course there are countless other alternatives, such as nonlinear trends or stochastic trends, which are just as plausible from the outset as a structural break in a linear trend. In the case of a smooth nonlinear trend, where the rate of increase changes at every point in time, the entire research question becomes meaningless and would have to be completely reformulated. Similarly, even if one assumes that ARMA models are suitable for describing the trend residuals, which in particular requires that the trend residuals are stationary and exhibit no seasonal patterns, the ARMA order $(p, q)$ must still be chosen. If this is not taken into account, too large rejection regions will be selected, thereby invalidating all test results. Indeed, pretending that the selected ARMA order is given rather than data-driven would raise the entire issue of post-model-selection inference (see, e.g., Pötscher, 1991).

If significance tests cannot provide the expected level of certainty, purely descriptive approaches should be given greater consideration. Just beefing up the raw data with a few simple transformations is enough to get a good sense of what is going on.

Figure 2 shows that a breach of the 1.5°C Paris target is imminent. This finding does not depend at all on when we set the beginning and end of the pre-industrial reference period, and in particular, it does not depend on whether or not we exclude the questionable data from before 1880. Similarly, Figure 3 shows a noticeable acceleration in the rate of warming. Simply by looking at the figure and noting the variability, we get a sense of the confidence level. It does not really matter where this variability comes from - whether from El Niño or some other source.

From a purist's point of view, a visual analysis has much less evidential value than a statistical test. But in our case, the results of more sophisticated investigations are very consistent with the visual findings. Using reasonably plausible assumptions, we were able to reject the linear-trend hypothesis (i.e., no acceleration) at the 10% level with relative ease. To reach the 5% level or even the 1% level, a few lucky choices helped - not all of which were strictly necessary. Also in the case of the second question, testing cannot reveal anything beyond what is already evident from the visual examination, namely that the breach of the 1.5°C Paris limit is imminent but has not yet occurred. At best, it might be possible to rule out a late breach (2028 or later) with a reasonable degree of certainty (Reschenhofer, 2025). But that, too, requires a number of assumptions regarding the reference period, the trend, and the deviations from the trend. However, given the approaching El Niño, we might be able to make more robust and reliable statements on this matter very soon.